\documentclass[manuscript, authorversion]{acmart}

\AtBeginDocument{%
  }

 \setcopyright{none}

\renewcommand\footnotetextcopyrightpermission[1]{%
  \begin{minipage}{\columnwidth}
  \textcolor{red}{\textit{*Note: Author's version. The final version will appear in EnCyCriS '26. DOI 10.1145/3786160.3788472}}
  \end{minipage}
}

\usepackage{hyperref}

\begin{document}

\title[]{Towards a Cognitive-Support Tool for Threat Hunters}

\author{Alessandra Maciel Paz Milani}
\affiliation{%
  \institution{University of Victoria}
  \city{Victoria}
  \country{Canada}
}
\orcid{https://orcid.org/0000-0001-8900-4179}
\email{amilani@uvic.ca}

\author{Norman Anderson}
\affiliation{%
  \institution{University of Victoria}
  \city{Victoria}
  \country{Canada}}
\orcid{https://orcid.org/0009-0003-1238-8014}  
\email{normananderson@uvic.ca}

\author{Margaret-Anne Storey}
\affiliation{%
  \institution{University of Victoria}
  \city{Victoria}
  \country{Canada}
}
\orcid{https://orcid.org/0000-0003-2278-2536}
\email{mstorey@uvic.ca}

\renewcommand{\shortauthors}{A. Maciel Paz Milani, N. Anderson, and M-A. Storey}

\begin{abstract}
Cybersecurity increasingly relies on threat hunters to proactively identify adversarial activity, yet the cognitive work underlying threat hunting remains underexplored or insufficiently supported by existing tools.
Building on prior studies that examined how threat hunters construct and share mental models during investigations, we derived a set of design propositions to support their cognitive and collaborative work.
In this paper, we present the \textit{Threat Hunter Board}, a prototype tool that operationalizes these design propositions by enabling threat hunters to externalize reasoning, organize investigative leads, and maintain continuity across sessions. 
Using a design science paradigm, we describe the solution design rationale and artifact development. 
In addition, we propose six design heuristics that form a solution-evaluation framework for assessing cognitive support in threat hunting tools. 
An initial evaluation using a cognitive walkthrough provides early evidence of feasibility, while future work will focus on user-based validation with professional threat hunters. 
\end{abstract}



\maketitle

\section{Introduction}
\label{intro}

The increasing sophistication and speed of cyber threats have made the work of threat hunters---security professionals who proactively search for undetected adversarial activity---vital and cognitively demanding. Threat hunting requires analysts to integrate fragmented data, form and revise hypotheses, and communicate evolving information across teams. While advances in automation and analytics continue to support detection, few tools are designed to explicitly support the human cognitive processes involved in understanding and reasoning about complex security events.

In our previous work~\cite{milani2025}, we conducted an in-depth observational study of professional threat hunters (THs) to examine how they build and refine their mental models during investigations. 
In this context, a mental model refers to the evolving internal representation THs form of an investigation, capturing relationships between events, hypotheses, entities, and expectations about attacker behavior and the client’s environment. 
From these findings, we proposed a set of five design propositions to improve cognitive support in threat hunting tools, emphasizing capabilities such as externalizing reasoning, maintaining continuity across sessions, and communicating investigative stories.

Building on those insights, this paper introduces the \textit{Threat Hunter Board}, a novel prototype tool designed to help THs externalize their mental models, track investigative leads, and collaborate more effectively with other THs and stakeholders. The tool introduces new interaction concepts, including \textit{waypoints} (markers of key discoveries or decisions), \textit{storylines} (persistent visual narratives of investigations), and \textit{checklists} (structured self-tracking for handovers). Together, these features embody a human-centered approach to supporting cognitive and collaborative work in cybersecurity analysis.

The primary contribution of this paper is the solution design of the \textit{Threat Hunter Board}, along with the presentation of our design science research process and an initial evaluation approach to assess its feasibility and perceived value—including a cognitive walkthrough and design heuristics.

\section{Background and Related Work}
\label{background}

In this section, we describe the process of \textit{threat hunting}, highlighting the myriad ways it is conducted, discuss tool use within the profession, and conclude by discussing considerations for designing tools to support threat hunters cognitively.

\textbf{Threat Hunting.} To threat hunt is to discover cyber threats which have evaded automated detection systems~\cite{bianco2023, nursidiq2024, nour2023_th_enterprise}. However, definitions for how this is accomplished range from ``any manual or machine-assisted process''~\cite{bianco2023} to an explicitly proactive and iterative approach~\cite{nursidiq2024, vanos2018}. In our work, we use the term \textit{threat hunter} (TH) to describe someone who threat hunts, in order to discuss how to best support the cognitive needs of these individuals. Security Operations Centers (SOCs) often encompass threat hunting, and THs may have another role within them or another organization, or work on a dedicated threat hunting team~\cite{badva2024}. This specialized role, the dedicated TH, has only recently emerged~\cite{milani2025}. Although we focus on human factors rather than the threat hunting process itself, we maintain that effective threat hunting must balance both of these considerations.

Although threat hunting as an activity has been discussed since the early 2010s~\cite{mahboubi_evolving_2024}, the field has key gaps in its procedural and organizational aspects~\cite{nour2023_th_enterprise}. 
For instance, in a recent industry report~\cite{lemon2025}, it is mentioned that threat hunting still lacks a formally accepted industry methodology for conducting hunts, despite several frameworks and methodologies defined in recent years (e.g., \cite{mitre, vanos2018, bianco2013, bianco2023}).
A similar lack of consensus exists regarding the ways in which a threat hunt may commence. Several existing taxonomies differ in their categorization of how a hunt begins~\cite{maxam2024, nour2023_th_enterprise}. Similarly, for the hunting process itself, although prior work commonly distinguishes between \textit{structured hunting} (or hypothesis method) and \textit{unstructured hunting}~\cite{nour2023_th_enterprise, maxam2024, vanos2018}, sometimes \textit{situational hunting} (or entity-driven) is distinguished as a third type of hunt~\cite{nour2023_th_enterprise}.

To summarize, threat hunting remains an individualized process with a variety of definitions and strategies and limited theoretical grounding. Tools designed to support THs must therefore account for a wide range of approaches and workflows.

\textbf{Cybersecurity Tools.} The tool landscape for threat hunting is diverse. Previous work has identified up to 95 distinct technical and non-technical tools THs used to support their work~\cite{hill2025}. We briefly discuss a smaller subset of industry tools which are identified by existing literature as designed to support THs in identifying threats~\cite{lemon2025, nour2023_th_enterprise}. Hunts making use of these tools are also informed by threat intelligence, such as knowledge of attackers' techniques, tactics, and procedures (TTPs)~\cite{vanos2018, badva2024}, and indicators of compromise (IoCs)~\cite{nour2023_th_enterprise, maxam2024, badva2024}.

According to Nour et al., enterprise security offerings ``overwhelmingly, provide threat detection capabilities rather than pursuing advanced threats and adversaries''~\cite{nour2023_th_enterprise}. They caution that the following solutions must \textit{support} rather than \textit{replace} threat hunting.
Security Information and Event Management (SIEM) tools are primarily designed to aggregate data from numerous systems. Security Orchestration, Automation, and Response (SOAR) tools improve upon SIEM via automating responses to incidents. eXtended Detection and Response (XDR) tools provide comprehensive monitoring with centralized detection and response to threats. User and entity behavior analytics (UEBA) methods can also detect threats, by comparing newly collected data against previously established baselines and anomalies as in need of further investigation~\cite{mahboubi_evolving_2024}.

Nour et al.~\cite{nour2023_th_enterprise} and Maboubi et al.~\cite{mahboubi_evolving_2024} taxonomize the academic approaches to tools aiding in the detection of threats.
Mahboubi et al.~\cite{mahboubi_evolving_2024} in their review additionally discuss tools associated with the MITRE ATT\&CK framework. However, the tool solutions in each of these categories are not designed to explicitly provide cognitive support during threat hunting.


\textbf{Cognitively Supporting Threat Hunters.}  
Previous literature has highlighted the lack of human factors research in threat hunting~\cite{hill2025, badva2024}.
This existing research has identified ways in which the working habits of THs lead to mental fatigue on the job (cognitive support needs) through an examination of how THs work, what tools they use, and what challenges they face.

For instance, mirroring the diversity of other aspects of threat hunting, THs disseminate the information they share to colleagues in a variety of ways: through handover or team meetings, asynchronously via message apps, presentations, and internal documentation~\cite{hill2025}. THs share in so many ways because of the number and heterogeneity of collaborative situations they encounter daily~\cite{hill2025}. Communication can be one-on-one or one-to-many; internal (teammates, managers, and across teams in the broader organization) or external (clients, cybersecurity insurance companies, supply chain vendors, and security researchers)~\cite{badva2024, hill2025}. In this context, it is unsurprising that information overload~\cite{hill2025, badva2024}, challenges simplifying and using information~\cite{hill2025, badva2024}, and constraints posed by context switching~\cite{hill2025} were all listed by THs as concerns.

Our recent work has continued to expand on existing threat hunting human factors research~\cite{milani2025}, introducing a model of how THs build mental models\footnote{See HCI research discussion on mental models in \cite{norman2013} and \cite{Hu2023}} during investigations and five design propositions to inform future tool development: 
(DP1) \textit{Creating a Story or Timeline of Events},
(DP2) \textit{Visualizing / Navigating Connections (Spatial)},
(DP3) \textit{Creatively Expanded Search},
(DP4) \textit{Waypoints and Note-Taking}, and
(DP5) \textit{Integrating External Resources}.

In the following section, we propose a solution to TH's cognitive support needs---one example of how the design propositions introduced in our prior work can be realized.

\begin{figure*}
  \includegraphics[width=1\linewidth]{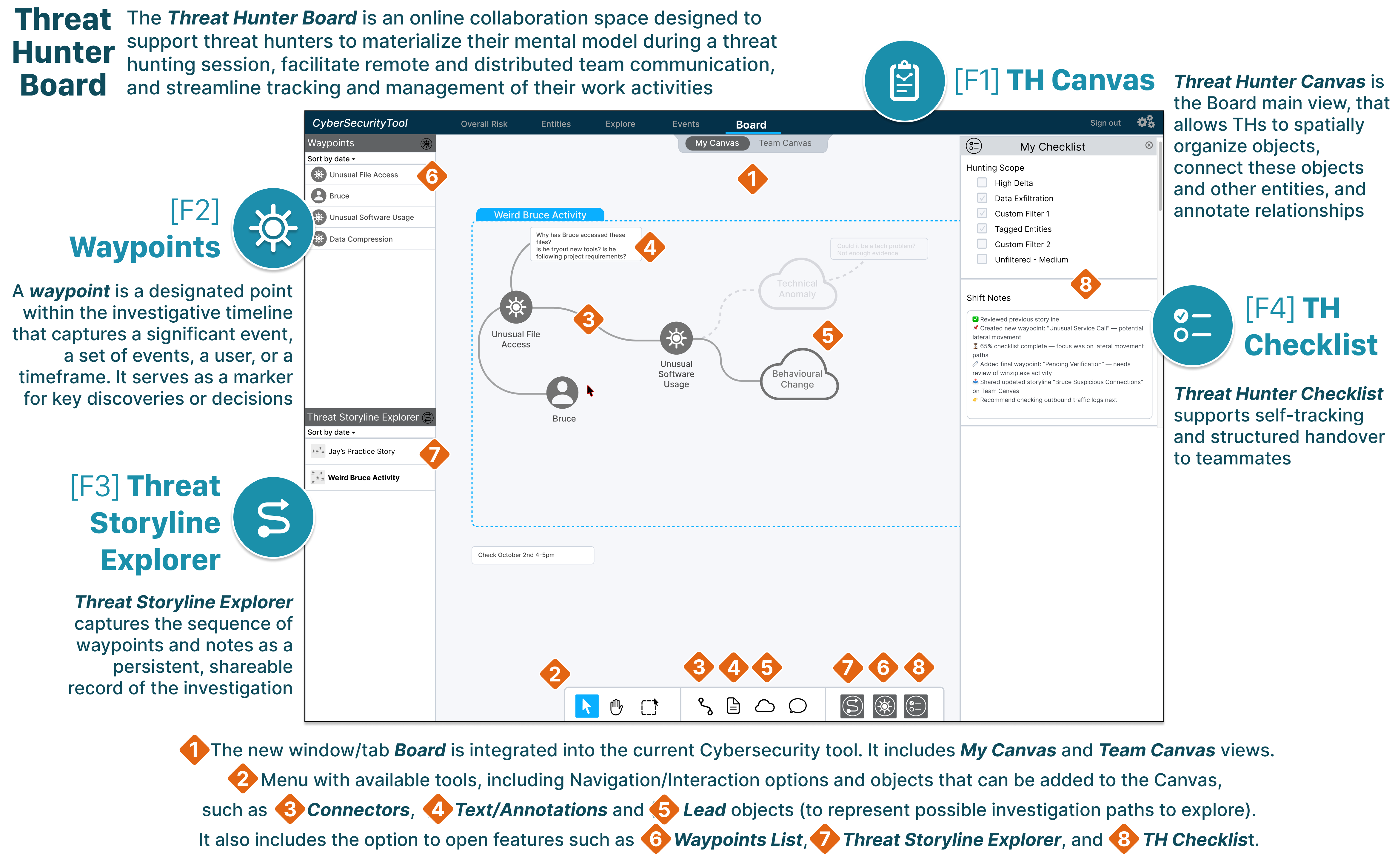}
  \caption{The \textit{Threat Hunter Board} and an overview of its core features.}
  \Description{Threat Hunter Board and an overview of its core features}
  \label{fig_overall_tool}
\end{figure*}

\section{Threat Hunter Board}
\label{sec_th_board_introduction}

Building on our prior findings, the \textit{Threat Hunter Board} was designed to embody the cognitive and collaborative needs identified in earlier work (see Sec.~\ref{background}). 
However, rather than restating the earlier design propositions, we refer to the conceptual structure guiding this prototype as a set of features. 
Each feature consolidates insights derived from:
(a) the cognitive model introduced in our previous work~\cite{milani2025};
(b) the design propositions defined there, 
and (c) related human-centered and cognitive frameworks explored in prior work.

We describe the research design approach in Section~\ref{methodology}; here, we introduce the \textit{Threat Hunter Board}. Figure~\ref{fig_overall_tool} provides an overview of the prototype and its main user interface elements.
We designed the \textit{Threat Hunter Board} prototype to integrate with existing cybersecurity environments. For example, by appearing as a new menu option within a current security solution. 

In our study scenario, we consider a TH primarily working with a UEBA tool (see Sec.~\ref{background}) and we assume that our target audience is familiar with such tools. Therefore, our focus in this paper is on presenting and discussing the novel design propositions and interaction concepts embodied in the \textit{Threat Hunter Board}, rather than on the underlying threat detection or analytics systems.

Next, we present the proposed features, and later, we illustrate their use through a scenario that highlights how the tool supports threat hunters during their investigative reasoning.\footnote{A video demo is available on \href{https://youtu.be/wP9DCgar__E}{https://youtu.be/wP9DCgar\_\_E}}

\subsection{Features}
\label{sec_features}

The \textit{Threat Hunter Board} is designed as an online collaboration space that helps THs materialize their mental model (that is, to externalize the evolving representation of the attack narrative that exists in their minds).
To operationalize this idea, we implemented four core features:
[F1]~\textit{Threat Hunter Canvas};
[F2]~\textit{Waypoints};
[F3]~\textit{Threat Storyline Explorer}; and
[F4]~\textit{Threat Hunter Checklist}.
These features are introduced in the next subsections, and a scenario illustrating their use is presented in Section~\ref{sec_th_scenario}.

\subsubsection{Threat Hunter Canvas}
The Threat Hunter Canvas represents the main view of the Board, where threat hunters can spatially organize investigation objects, connect entities, and annotate relationships. The design goal was to make the interface as simple and intuitive as a notepad, yet flexible enough to support free-form exploration similar to sketching on paper (drawing inspiration from collaborative design tools such as Miro\footnote{\href{https://www.miro.com/}{https://www.miro.com/}} and Figma\footnote{\href{https://www.figma.com/}{https://www.figma.com/}}).

The Canvas provides two view modes: \textit{My Canvas}, which supports individual exploration, and \textit{Team Canvas}, which facilitates shared understanding and collaboration among multiple analysts. A contextual menu provides access to available tools and interaction options, including navigation controls and object creation such as \textit{Connectors}, \textit{Text/Annotations}, and \textit{Lead objects} (which represent potential investigative paths to be explored).

Additionally, Canvas integrates other key components of our proposed tool: \textit{Waypoints List}, \textit{Threat Storyline Explorer}, and \textit{Threat Hunter Checklist}, each of which is described in detail next.

\subsubsection{Waypoints}
A \textit{Waypoint} represents a designated point within the investigative timeline that captures a significant event, a group of related events, a specific user, or a defined timeframe. Each waypoint acts as a marker for key discoveries or analytical decisions made during the investigation.

The waypoint concept can be implemented in various forms, depending on its integration with existing threat hunting tools. In our prototype, developed to integrate with the same tool used in our previous observational study conducted with our industry partner (see Sec.~\ref{sec_industry_partner}), a new \textit{Waypoint} icon can be introduced alongside the existing \textit{Bookmark} icon. Similar to a bookmark that saves a system view (e.g., the URL with the parameters of the current query), a waypoint extends this functionality by storing additional metadata relevant to the investigation context (such as waypoint name, notes, type, details, and event period---most of the fields are pre-filed).

Once created, the waypoint appears in the \textit{Waypoint List}, where users can manage, filter, and sort saved waypoints. This list is accessible from the Board interface as a floating panel positioned on the left side of the canvas. 
Waypoints can be dragged and dropped onto the canvas to visually associate them with specific investigative stories or analytical threads (covered in the next feature).

In this implementation, waypoints are displayed as rounded objects, each with a distinct symbol at its center indicating its type.
In this example (see Figure~\ref{fig_overall_tool}), a headlight-like icon represents an event, while a person-shaped icon denotes a user. Colors can also be applied to waypoints to indicate their priority level (e.g., highlighting high-priority items). The associated metadata for each waypoint can be viewed or edited either from the \textit{Waypoint List} or directly on the Canvas, where it appears as a floating menu or in a dedicated right-hand panel.

\begin{figure}
  \includegraphics[width=\columnwidth]{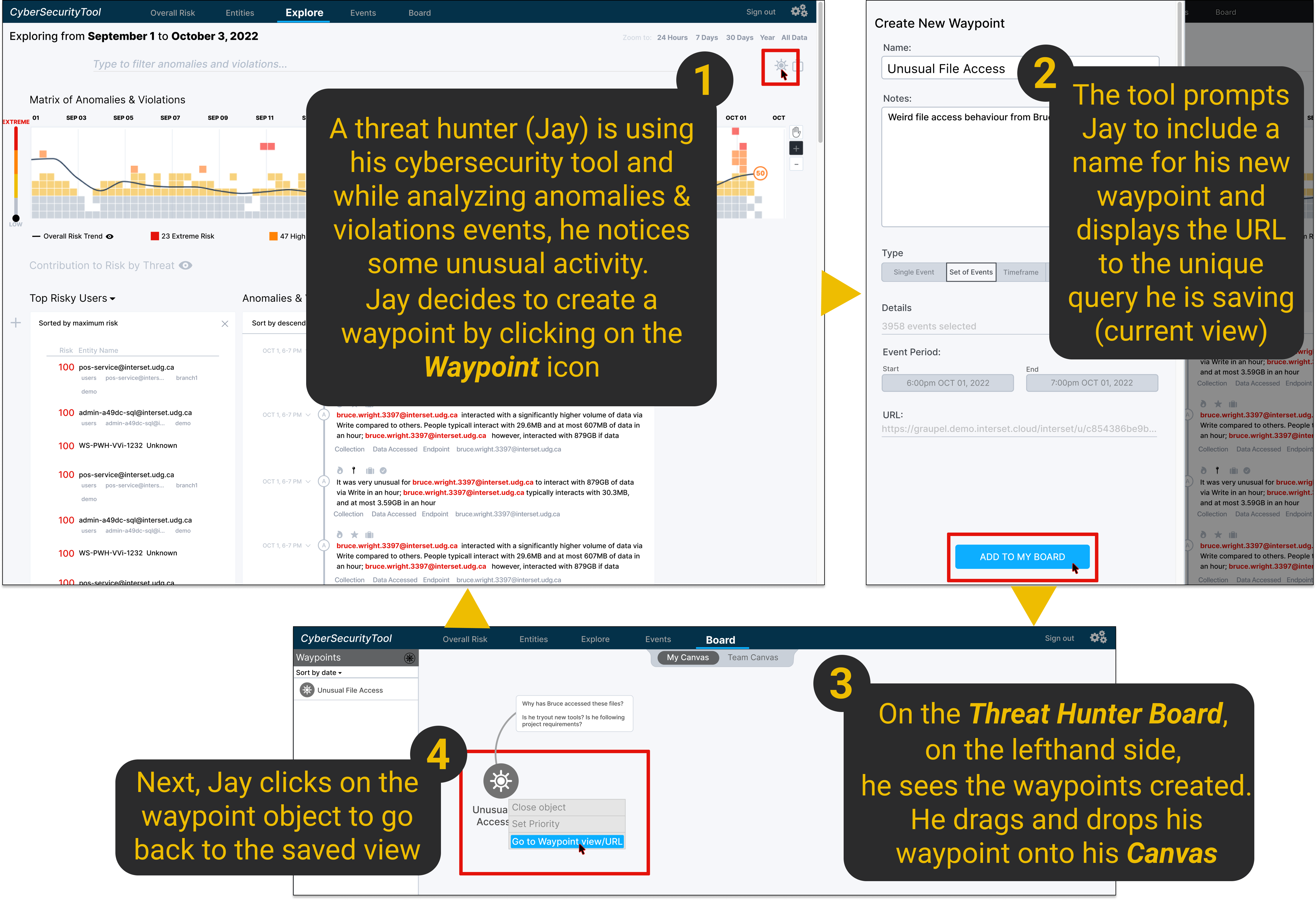}
  \caption{F2 Waypoints---example of a waypoint creation.}
  \Description{Example of a waypoint creation}
  \label{fig_waypoint_creation}
\end{figure}

\begin{figure}
  \includegraphics[width=\columnwidth]{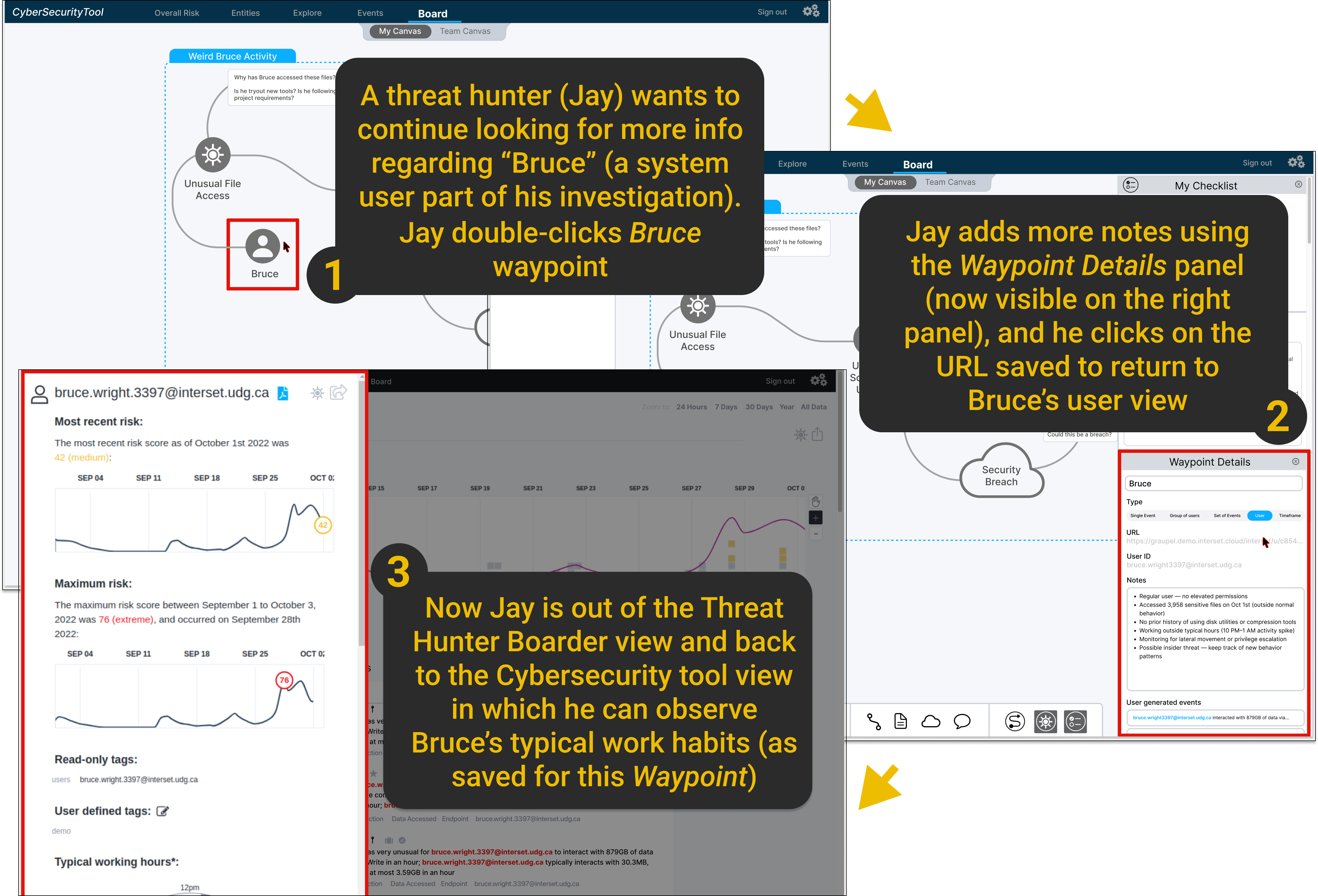}
  \caption{F2 Waipoints---example of a waypoint object being edited, and opening its saved view.}
  \Description{Example of a waypoint object being edited, and opening its saved view}
  \label{fig_waypoint_opening}
\end{figure}

\subsubsection{Threat Storyline Explorer}

Within the Canvas, THs can group and save a selected set of objects, such as waypoints, leads, and text annotations, as a new \textit{Threat Storyline}. This \textit{storyline} represents a coherent segment of the investigation that can be shared and reviewed by peers through the \textit{Team Canvas}, supporting collaborative sense-making and knowledge continuity. The most recently created storyline is automatically displayed in the user's \textit{My Canvas} view, while previously saved storylines can be accessed from the panel on the left side of the Board. 

This design supports seamless transitions between individual and collaborative work, enabling analysts to revisit, extend, or refine prior investigations without losing contextual information.

\subsubsection{Threat Hunter Checklist}

The \textit{Threat Hunter Checklist} is designed to support THs while self-tracking, session planning, and to support structured handovers across team members. It provides a lightweight, yet structured, mechanism for THs to ensure coverage of key tasks throughout their shift.

Each checklist instance is associated with a specific client environment, reflecting the common setup in which each cybersecurity tool deployment is dedicated to a single client organization (as observed in our previous study).

Checklist items can be created from a predefined template aligned with standard threat hunting procedures but can also be customized to reflect the TH's unique workflow or investigative goals. As the session progresses, the checklist enables THs to track completed and pending activities, providing an overview of what has been covered and what remains to be explored.

With this feature, the \textit{Threat Hunter Board} also supports session continuity by allowing the TH to save their current state as a bookmark linked to the checklist, making it easier to resume from the exact point where the investigation was paused. Finally, completed checklists and notes can be shared through the \textit{Global Checklist} or \textit{Handover Board} (possible new features to be continued), ensuring that key findings, progress updates, and pending tasks are clearly communicated to teammates in subsequent shifts.

The \textit{Threat Hunter Checklist} thus reinforces both individual cognitive support, by helping THs manage their workload and focus, and team-level coordination, by providing a structured process for shift transitions and collective situational awareness.

\subsection{Showcase Scenario: Illustrating the Threat Hunter Board in Use}
\label{sec_th_scenario}

To illustrate the design and features of the \textit{Threat Hunter Board}, we present a scenario that demonstrates how the tool supports a TH's cognitive process during an investigation. The following vignette features \textit{Jay} (TH persona from~\cite{hill2025}), an experienced external consultant who conducts proactive threat hunting sessions for multiple clients. 
The scenario is inspired by common workflows and tool interactions observed in our prior study~\cite{milani2025}, which revealed how THs iteratively construct and externalize their mental models as they form and test hypotheses.
Additionally, the dataset and overall investigation narrative were adapted from a threat hunting training exercise provided by our industry partner. While not based on real organizational data for privacy and confidentiality reasons, the scenario reflects realistic threat hunting situations observed in practice (i.e., a based-on-true-events representation of actual user interface and investigative workflows).

\textbf{[Scene 1] Starting from Scratch.}
At the beginning of his shift, Jay opens the \textit{Threat Hunter Board} and reviews his team's previous notes under the \textit{TH Checklist}. In his client’s security tool, he notices an anomaly: a user named ``Bruce Wright'' has accessed more than three thousand files in a single session. Recognizing this as unusual, Jay creates his first \textit{waypoint} labeled ``Unusual File Access,'' which automatically saves the corresponding query from his analytics tool (see Figure~\ref{fig_waypoint_creation} for details). Switching to the \textit{Board} view, Jay drags the waypoint onto his \textit{Canvas} and adds a note hypothesizing that Bruce’s activity may relate to data migration.

Upon inspecting Bruce's user permissions, Jay rules out this hypothesis and creates a second waypoint titled ``Bruce'' (User Entity). On the Board, he links these two waypoints with a \textit{connector}, visualizing the relationship between the event and actor. He then saves his progress as a \textit{Threat Storyline} titled ``Weird Bruce Activity.''

\textbf{[Scene 2] Exploring Multiple Paths.}
A few days later, in another session, Jay detects a new anomaly: Bruce has used an unusual software executable to create disk partitions. Jay creates another waypoint (``Unusual Software Usage'') and loads the previous storyline ``Weird Bruce Activity''. On the canvas, he connects this new waypoint to his earlier ones, revealing a growing chain of events.

\begin{figure}
  \includegraphics[width=\columnwidth]{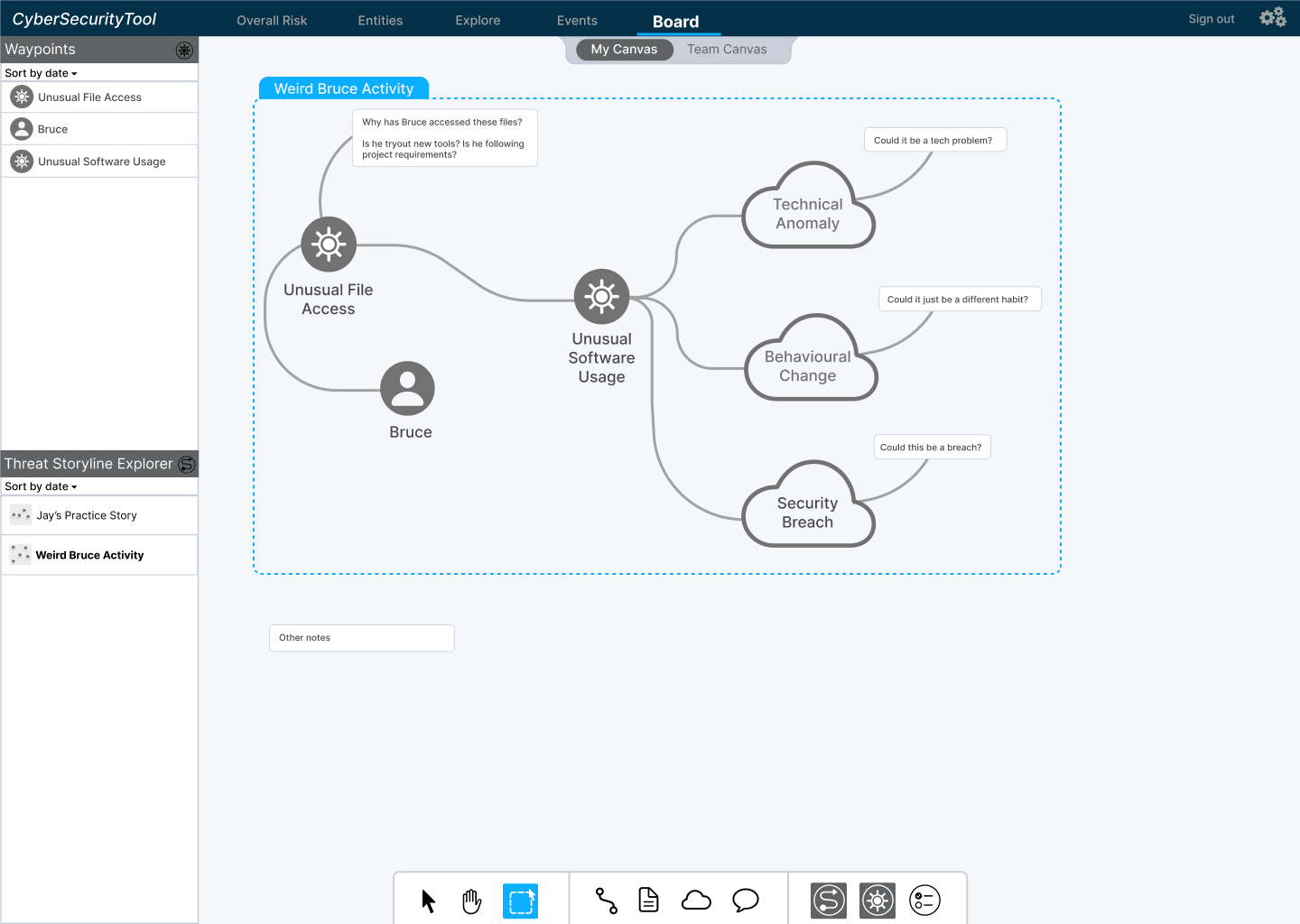}
  \caption{Showcase scenario---view of the \textit{Threat Hunter Board} at the end of \textit{Scene 2}.}
  \Description{Showcase scenario view of the Threat Hunter Board at the end of Scene 2}
  \label{fig_scene2}
\end{figure}

Jay identifies three possible investigative paths: (1) Technical Anomaly, (2) Behavioral Change, and (3) Security Breach. He represents these hypotheses as \textit{lead} objects linked to the latest waypoint. Jay \textit{annotates} each lead with questions and notes, using the Board to externalize his reasoning process. This visual structure enables Jay to maintain a clear mental model while keeping multiple hypotheses active (see Figure~\ref{fig_scene2} for details).

\textbf{[Scene 3] Narrowing Down the Options.}
As Jay continues investigating, he discovers that Bruce has used ``winzip.exe'' to compress the files associated with the previous partitioning event (see Figure~\ref{fig_waypoint_opening} for details on how the waypoint ``Bruce'' was accessed and edited from the canvas). He creates a new waypoint titled ``Data Compression'' and links it to the relevant leads. After reviewing the visual relationships between waypoints, Jay concludes that the ``Technical Anomaly'' hypothesis is no longer viable and marks that \textit{lead} as closed. The canvas now reflects his refined mental model: two active hypotheses---``Behavioral Change'' and ``Security Breach''---supported by a chain of evidence.

\begin{figure}
  \includegraphics[width=\columnwidth]{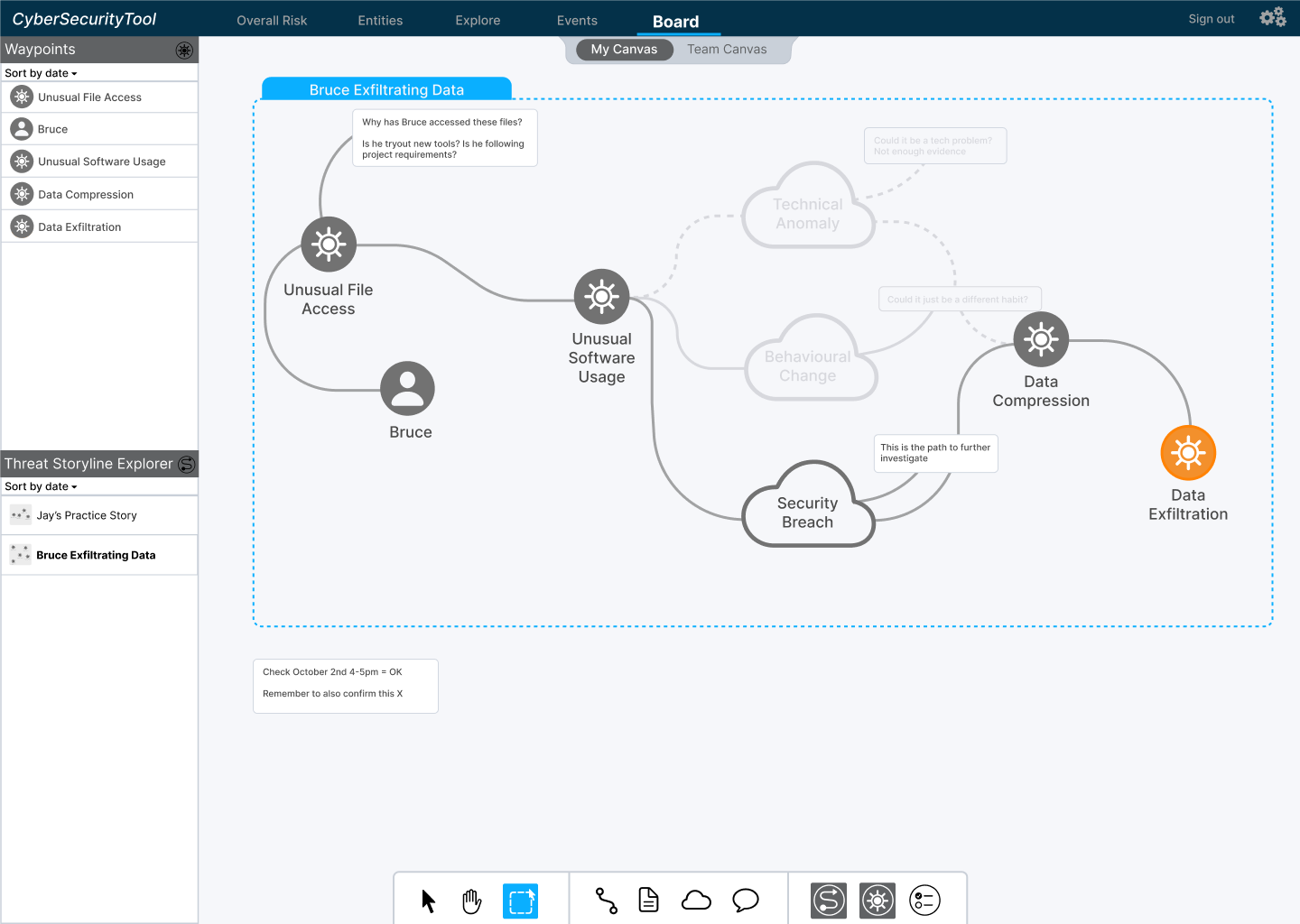}
  \caption{Showcase scenario---view of the \textit{Threat Hunter Board} at the end of \textit{Scene 4}.}
  \Description{Showcase scenario view of the Threat Hunter Board at the end of Scene 4}
  \label{fig_scene4}
\end{figure}

\textbf{[Scene 4] Confirming the Suspicion.}
In a later session, Jay receives an alert showing that Bruce has uploaded over 800 GB of data to an external volume. Jay creates a final waypoint labeled ``Data Exfiltration,'' marks it as high priority, and links it to previous data-related events. The visual map now provides a coherent narrative from ``Unusual File Access'' to ``Data Exfiltration.'' Jay closes the remaining ``Behavioral Change'' lead and renames the \textit{threat storyline} ``Bruce Exfiltrating Data.'' His externalized model now represents a clear hypothesis confirmed by converging evidence (see Figure~\ref{fig_scene4} for details).

\textbf{[Scene 5] Collaboration and Handover.}
Before ending his shift, Jay opens the \textit{Team Canvas} view, where he shares the ``Bruce Exfiltrating Data storyline'' with his colleagues. Other threat hunters can now review the visualized investigation, access saved queries, and continue exploring open questions. Using the \textit{integrated checklist}, Jay verifies that all procedural items are complete and adds final notes for his handover meeting.
This collaborative workflow allows the next analyst to resume the investigation seamlessly, with access to Jay's full cognitive trace (his reasoning, hypotheses, and evidence) preserved in the storyline.

\textbf{Summary.} 
This showcase scenario demonstrates how the \textit{Threat Hunter Board} operationalizes core design principles derived from our earlier research. By enabling threat hunters to externalize their thought process through waypoints, visualize evolving hypotheses, and structure collaboration through shared checklists and storylines, the tool helps make the \textit{fuzzy-to-clear} cognitive transition more explicit and persistent.

\section{Research Approach}
\label{methodology}

Our research follows a design science paradigm~\cite{Runeson2020}, which emphasizes the creation and iterative refinement of an artifact to address a well-defined practical problem while contributing to theoretical understanding. In our case, the artifact---the \textit{Threat Hunter Board}---was developed to support threat hunters in externalizing and sharing their mental models during complex investigations. 

\begin{figure*}
  \includegraphics[width=1\linewidth]{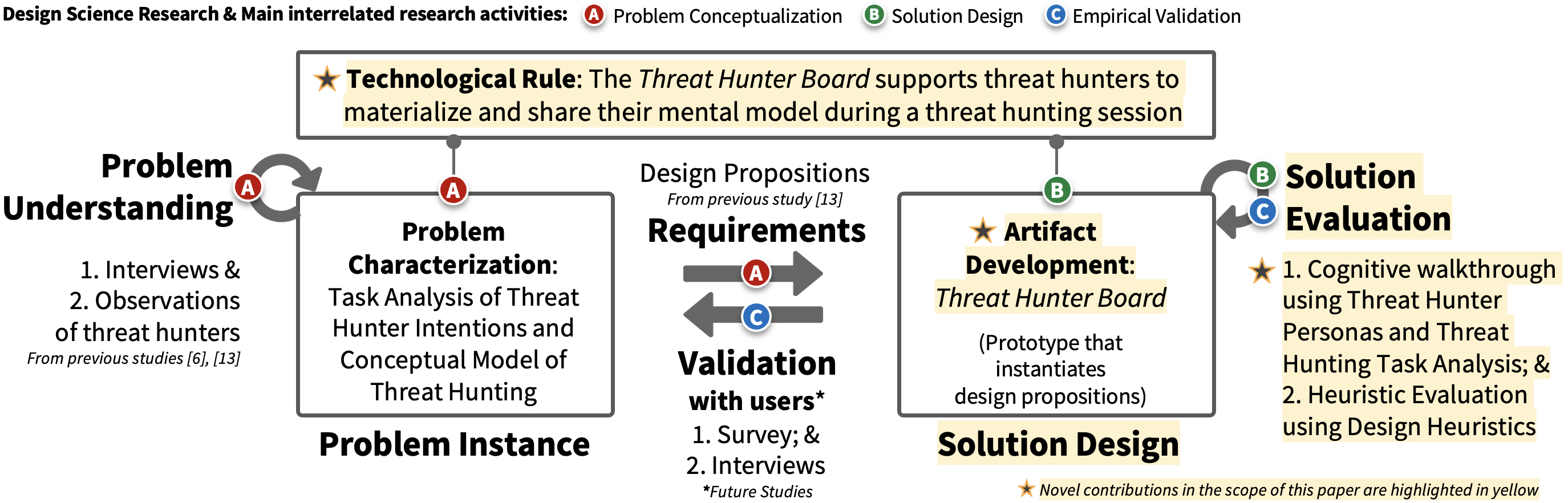}
  \caption{Our overall design science process. The parts shown in yellow are the main contributions shared in this paper.}
  \Description{Overall design science process}
  \label{fig_design_overview}
\end{figure*}

Figure~\ref{fig_design_overview} illustrates our overall design science research, and in the following subsection, we provide further details of this process.

\subsection{Design Science Process}
\label{sec_design_process}

The design science research process adopted in this work can be summarized through three interrelated activities: problem conceptualization, solution design, and empirical validation (as in \cite{Runeson2020}). We briefly explain each of them below. 

(A) \textbf{Problem Conceptualization}. 
Our research builds on the theoretical and empirical foundations established in prior work~\cite{hill2025,milani2025}. In particular, our most recent study~\cite{milani2025} investigated the cognitive processes of threat hunters through in-depth observations of real-world threat hunting activities. This work resulted in a set of design propositions that characterize key cognitive challenges and desired effects, which serve as guiding requirements for the design of cognitively supportive tools.

(B) \textbf{Solution Design}.    
Grounded in this problem understanding, we developed the \textit{Threat Hunter Board} as a proof-of-concept artifact. The solution design leverages a concrete problem instance to clarify the core practical challenges and operationalizes selected design propositions into an interactive prototype (see Sec.~\ref{sec_design_solution}). 

(C) \textbf{Empirical Validation}. 
To provide initial evidence of feasibility, we conducted a preliminary evaluation of the solution using two complementary methods. First, we performed a cognitive walkthrough based on TH personas and task analyses derived from earlier studies, assessing the alignment of the tool with realistic cognitive workflows (see Sec.~\ref{sec_th_scenario}). 
Second, we designed a heuristic evaluation framework using design heuristics for human-centered systems adapted to the cybersecurity domain (see Sec.~\ref{sec_design_process_heuristics}).

Consistent with design science principles, this evaluation represents an initial validation step. Future work will focus on validating the artifact with professional THs (see Sec.~\ref{sec_design_process_validation}). 
Nonetheless, we have received early qualitative feedback from key stakeholders at OpenText (our industry partner, see Sec.~\ref{sec_industry_partner}), which reinforces our designed solution's potential to provide cognitive support during threat hunting sessions.

In line with the design science paradigm~\cite{Runeson2020}, our work contributes both a concrete artifact and transferable design knowledge. Beyond the \textit{Threat Hunter Board} prototype, our research distills insights into how cognitive-support principles can be operationalized in cybersecurity tools, thereby linking human cognition theory with practical design decisions. This contribution supports rigor through empirically grounded design choices and relevance by addressing real-world challenges faced by professional THs.

\subsection{Industry Partner}
\label{sec_industry_partner}
Our research project is conducted in collaboration with OpenText.\footnote{https://www.opentext.com/products/cyber-security}
OpenText is a global market leader in information management software offering a variety of cybersecurity products and services. Among its offerings is an advanced threat detection platform that leverages user and entity behavior analytics (UEBA) and machine learning models to identify behavioral anomalies within organizational environments.\footnote{\href{https://www.opentext.com/products/arcsight-intelligence}{https://www.opentext.com/products/arcsight-intelligence}}

While our research is independently led by the university research team, representatives from OpenText contributed to specific stages of the research project, helping with participant recruitment and will support further validation of our research findings with their in-house cybersecurity experts on the proposed solution presented here---\textit{Threat Hunter Board}.

\subsection{Artifact Development}
\label{sec_design_solution}

The \textit{Threat Hunter Board} was implemented as a high-fidelity prototype using Figma. Although not fully interactive (i.e., not connected to a real back-end system), the design was intentionally developed to closely approximate real user experiences and workflows observed in professional threat hunting practice (i.e., it mirrors the layout of our industry partner's existing tool).

Our prototype instantiates design propositions (DPs) introduced in previous work~\cite{milani2025}, particularly \textit{Creating a Story or Timeline of Events} (DP1) and \textit{Waypoints and Note-Taking} (DP4). In this iteration, we also introduce a new feature to support an end-to-end threat hunting investigation session---the \textit{Threat Hunter Checklist}---not explicitly described in earlier DPs but reflected in our prior findings (i.e., the thematic data analysis of our observations also reported in \cite{milani2025}). To capture this addition, we extend our set of DPs with \textit{Threat Hunter Profile} (DP6) and \textit{Checklists (Individual and Global for Handover)} (DP7). Although detailed user stories for the new DP6 and DP7 are not discussed here due to space constraints, they represent another relevant direction emerging from this phase of our research.

As a final remark, our prototype development and scope follow a \textit{minimum viable product} and \textit{iterative design} approach. It is not intended to be exhaustive in its implementation; the features described (Sec.~\ref{sec_features}) can be further refined, and additional functionalities may be developed to meet the already identified problems and the evolving user needs.

\subsection{Design Heuristics}
\label{sec_design_process_heuristics}

To support a systematic evaluation of the proposed solution, we define six interrelated design heuristics (DHs) that together form the basis of our solution evaluation framework. Each DH captures a specific dimension of cognitive support relevant to threat hunting and articulates intended interaction qualities and design outcomes. 
The heuristics are grounded in empirical findings from prior studies and established theoretical perspectives (e.g., HCI fundamentals~\cite{mackenzie_hci_2024} and technology acceptance models \cite{davis_tam_2024}), and are used to assess how effectively the \textit{Threat Hunter Board} addresses key cognitive and collaborative challenges.

(DH1) \textbf{Navigation \& Exploration}. Supports users in traversing related entities, events, or data while maintaining orientation and the ability to retrace investigative paths.

(DH2) \textbf{Clarity \& Sense-Making}. Helps organize fragmented or ambiguous data into coherent investigative narratives that clarify evolving hypotheses.

(DH3) \textbf{Decision Support}. Provides structure and cues to assist threat hunters in prioritizing, shifting focus, or determining when to conclude an investigation.

(DH4) \textbf{Communication \& Handover}. Facilitates the transfer of reasoning and evidence across individuals or teams, enabling continuity and collaborative sense-making.

(DH5) \textbf{Memory \& Mental Load}. Reduces reliance on short-term memory by externalizing information, tracking progress, and resuming suspended work.

(DH6) \textbf{Overall Perception}. Reflects the holistic experience of usefulness, usability, and the prototype's perceived fit within existing workflows and tools.

These six DHs enable a structured assessment of the problem solution instance by making cognitive support requirements explicit and evaluable. Beyond their use in our research, the DHs presented here are intended to be reusable by other tool designers to conduct heuristic evaluations of similar cybersecurity tools and to serve as a foundation for future user-based validation with professional THs (see Sec.~\ref{sec_design_process_validation}).
As such, the DHs constitute a core design knowledge contribution of this paper.

\subsection{User-based Validation Plan}
\label{sec_design_process_validation}

Our future study will follow a mixed method approach integrating both quantitative and qualitative data collection, in a convergent parallel design \cite{storey_guiding_2025}. 
Next, we explain the planned data collection instruments, target participants and their recruitment strategy, and we close with the data analysis plan.

\textbf{Data Collection Instrument}.
An online survey and interviews are planned to be conducted in parallel for data collection. 
Each participant session will begin with a brief introduction to the research objectives and confirmation of informed consent, followed by:
(a) viewing a prototype demonstration video;
(b) completing closed-ended evaluation questions aligned with the six DHs; and
(c) participating in open-ended discussions to elaborate on their ratings, challenges, and expectations regarding cognitive-support tools.
The main distinction between the two methods lies in their depth and reach. While the survey enables broader participation, the interviews allow for deeper validation through follow-up questions and richer qualitative insights. 
See supplementary materials for the data collection instruments \cite{milani_2025_17511203}.

\textbf{Target Participants and Recruitment}. Participants will be recruited and invited through professional and academic cybersecurity networks. Eligible participants include individuals with direct experience in threat hunting---either as a dedicated role or as a recurring component of their cybersecurity work. 

\textbf{Data Analysis}. 
For the survey, the quantitative responses (Likert-scale ratings) will be analyzed descriptively to identify perceived strengths and areas for improvement across the six DHs.
Qualitative data from open-ended survey items and interview transcripts will be analyzed using thematic analysis~\cite{braun_thematic_2022}. 
Codes will be first derived inductively from participant responses, then mapped to the six DHs to reveal recurring perceptions and cognitive support needs.
This dual analysis will triangulate participant feedback across structured and exploratory data sources.

While user-based validation is outside the scope of this paper, we present the plan here to demonstrate how the proposed DHs can be used and to provide a clear path for future empirical validation of the \textit{Threat Hunter Board}.

\section{Discussion}
\label{discussion}

The \textit{Threat Hunter Board} exemplifies how cognitive-support principles can be operationalized through tool design in cybersecurity analysis.
By translating the design propositions derived from our previous work~\cite{milani2025} into tangible interaction concepts, our prototype demonstrates how visual externalization, structured tracking, and collaborative storyline can reduce cognitive overhead and enhance continuity in complex investigations.

\textbf{Supporting Cognitive Processes.}
The solution design directly targets key cognitive mechanisms identified in earlier work: mental model externalization, sense-making, and memory offloading.
Features such as \textit{Waypoints} and the \textit{Threat Storyline Explorer} allow THs to capture and visualize evolving hypotheses, making implicit reasoning explicit and persistent.
Similarly, the \textit{Threat Hunter Checklist} encourages self-monitoring and continuity across shifts, providing structured cues that reduce the cognitive cost of context switching.
These mechanisms align with prior findings that THs often rely on scattered notes or ad-hoc mental tracking, which increases cognitive load and impedes collaboration.

\textbf{Human-Centered Integration.}
Unlike many recent efforts that focus on automating threat detection or applying machine learning to identify anomalies, our work emphasizes the human-in-the-loop aspect of cybersecurity operations.
Rather than replacing analytical reasoning, the prototype complements existing UEBA or SIEM systems by giving analysts cognitive and organizational tools to interpret and act on machine-generated insights.
This reinforces the value of human-centered tool design, where automation and cognition coexist in a mutually supportive workflow.

\textbf{Methodological Reflection.}
Developing and evaluating the prototype through a \textit{design science research} process provided both practical and theoretical contributions.
Iterative design cycles allowed us to move from problem characterization to artifact realization while maintaining traceability between theoretical constructs (design propositions) and implemented features.
The six design heuristics---\textit{Navigation \& Exploration}, \textit{Clarity \& Sense-Making}, Decision Support, \textit{Communication \& Handover}, \textit{Memory \& Mental Load}, and \textit{Overall Perception}---offer a reusable framework for assessing future cognitive-support tools in similar domains.

\textbf{Comparison with Existing Approaches.}
Existing cybersecurity tools and related work largely emphasize data aggregation, automation, or procedural guidance for threat hunting. In parallel, lightweight artifacts such as checklists or community-developed templates support task tracking and documentation (e.g., \textit{Cybersecurity Threat Hunting Dashboard}\footnote{\href{https://www.notion.com/templates/cybersecurity-threat-hunting-dashboard}{https://www.notion.com/templates/cybersecurity-threat-hunting-dashboard}}). While useful, these approaches typically remain detached from THs' investigative context and provide limited support for externalizing and evolving mental models during active hunts.
In contrast, the \textit{Threat Hunter Board} is designed as a cognitive workspace integrated with existing threat hunting tools rather than as a standalone template or detection mechanism. Its contribution lies in supporting how THs reason, reflect, and communicate by enabling spatial externalization of hypotheses, evidence, and decisions while maintaining links to underlying data, queries, and investigative actions. This integration distinguishes our approach from static templates and generic collaboration tools, which lack direct coupling to THs' sense-making processes and system state.

\textbf{Limitations.} As a design-oriented and exploratory endeavour, our work has several limitations.
First, the \textit{Threat Hunter Board} remains a high-fidelity static prototype developed to demonstrate design concepts rather than a fully interactive system.
As such, the evaluation criteria focus on perceived usefulness and cognitive alignment rather than performance outcomes or task efficiency.
Second, the current user-based validation plan, based on surveys and interviews, targets a limited sample of expert practitioners, which may not represent the full diversity of threat hunting practices across organizations or tool ecosystems.
Additionally, our research context and collaboration with a single industry partner influence the design integration of the prototype, particularly its alignment with UEBA-style platforms.

\textbf{Future Work.} As the next step, we plan to conduct the validation studies with TH professionals (as explained in Sec.~\ref{sec_design_process_validation}). However, other future work can also:
(a) implement an interactive version of the prototype integrated with real threat hunting datasets;
(b) conduct longitudinal evaluations to assess how cognitive-support features affect workflow continuity and collaboration over time; and
(c) explore extensions of the \textit{Threat Hunter Board} to examine how it can interact with AI-assisted reasoning and cognitive analytics frameworks, thereby bridging human cognitive processes with intelligent automation to support next-generation threat hunting.

\section{Final Considerations}

In this paper, we translate prior insights on cognitive support for threat hunting into a concrete artifact and an accompanying evaluation framework. 
Through the solution design of the \textit{Threat Hunter Board}, we show how empirically grounded design propositions can be instantiated into features that support externalize reasoning, reduce cognitive load, and strengthen collaboration during complex threat hunting investigations. 
Rather than enforcing a fixed workflow, the \textit{Threat Hunter Board} supports diverse investigations while enabling reusable patterns (e.g., waypoints, storylines, or checklist elements) that could facilitate knowledge sharing across threat types or teams.
Beyond the prototype, this work contributes transferable design knowledge and demonstrates how a design science research approach can connect empirical understanding with human-centered cybersecurity tool design.

\begin{acks}
The authors would like to thank and report the financial support provided for project research during this study by Mitacs Canada, OpenText Corporation, and the Natural Sciences and Engineering Research Council of Canada (NSERC).
The authors also thank the CHISEL members who contributed to brainstorming and refining the solution design presented in this paper (especially Samantha Hill, David Moreno-Lumbreras, and Arty Starr).
\\
\textcolor{red}{\textit{*Note: Author's version. The final version will appear in EnCyCriS '26. DOI 10.1145/3786160.3788472}}
\end{acks}

\bibliographystyle{ACM-Reference-Format}
\bibliography{bibliography}

\end{document}